# Three-dimensional Epanechnikov mixture regression in image coding

Boning Liu[a], Yan Zhao[a,*], Xiaomeng Jiang[b], Shigang Wang[a]

[a] *College of Communication Engineering, Jilin University, Changchun 130012, P. R. China*
[b] *College of Mathematics, Jilin University, Changchun 130012, P. R. China*

## ABSTRACT

Kernel methods have been studied extensively in recent years. We propose a three-dimensional (3-D) Epanechnikov Mixture Regression (EMR) based on our Epanechnikov Kernel (EK) and realize a complete framework for image coding. In our research, we deduce the covariance-matrix form of 3-D Epanechnikov kernels and their correlated statistics to obtain the Epanechnikov mixture models. To apply our theories to image coding, we propose the 3-D EMR which can better model an image in smaller blocks compared with the conventional Gaussian Mixture Regression (GMR). The regressions are all based on our improved Expectation-Maximization (EM) algorithm with mean square error optimization. Finally, we design an Adaptive Mode Selection (AMS) algorithm to realize the best model pattern combination for coding. Our recovered image has clear outlines and superior coding efficiency compared to JPEG below 0.25bpp. Our work realizes an unprecedented theory application by: (1) enriching the theory of Epanechnikov kernel, (2) improving the EM algorithm using MSE optimization, (3) exploiting the EMR and its application in image coding, and (4) AMS optimal modeling combined with Gaussian and Epanechnikov kernel.

## 1. Introduction

Kernel methods have been widely applied in the artificial intelligence field such as computer vision [1] and pattern recognition [2]. The most common function in machine learning is the *Gaussian Kernel* (GK) [3], and the kernel methods provide tremendous application potential in other kernel functions. For example, correntropy has different properties when compared with second-order statistics that can be very useful in non-Gaussian signal processing [4]. Graph signals can also be recovered through a kernel-based reconstruction of space-time function [5]. In the context of deep learning, using deep architectures to perform kernel machine optimization has been explored [6]. Recently, studies on using kernel methods have been extended for developing non-parametric kernels. A regression approach was proposed to extend the non-parametric kernel matrix to the corresponding kernel function [7].

In such a trend, research studies in image coding are also expanding to kernel regression frameworks. Among them, the most notable was the *Steered Mixture-of-Experts* (SMoE), a model-based approach proposed by R. Verhack [8]. In this model, it was assumed that image pixel values are instantiations of a non-linear random process that can be modelled by spatially piecewise stationary Gaussian processes. SMoE can be flexibly applied to the image and light field representation, processing, and coding [9]. Apart from 2-D image coding, SMoE has also been applied to 4-D light field images and 5-D light field video representation [10,11], which fully demonstrates the superiority of model-based coding in higher dimensions. Moreover, several studies were conducted to optimize the performance of SMoE. In [12], a Gradient Descent optimization using the *Mean Square Error* (MSE) of the regressed imagery was proposed. Based on MSE, the Gradient Descent optimization for the efficiency of different covariance representations has also been discussed in [13]. It has been proven that the SMoE regressed by SSIM loss can compete with JPEG2000 even for high bitrates [14]. After all, SMoE is a revolutionary approach for image compression that drastically departs from traditional pixel-based coding algorithms. However, a disadvantage of SMoE is the limitation toward kernel types. Therefore, we aim to propose a new kernel theory and its application in image compression by modeling.

The *Epanechnikov Kernel* (EK) is an excellent kernel function owing to its computational convenience and concentration of distribution [15]. This kernel function has achieved promising results in many machine learning applications. In a related work, 1-D

☆ This work is supported by the National Natural Science Foundation of China (No.61631009, No.61771220) and the National Key R&D Program of China (2017YFB1002900, 2017YFB0404800).
* Corresponding author.
*E-mail addresses:* liuboning_jlu@sina.com (B. Liu), zhao_y@jlu.edu.cn (Y. Zhao), jxmlucy@sina.com (X. Jiang), wangshigang@vip.sina.com (S. Wang).

https://doi.org/10.1016/j.sigpro.2021.108090
0165-1684/© 2021 The Authors. Published by Elsevier B.V. This is an open access article under the CC BY-NC-ND license (http://creativecommons.org/licenses/by-nc-nd/4.0/)

Epanechnikov kernels were used to calculate the expected value of the irradiance estimators in photon mapping density estimation [16]. The EK can estimate function density in anomaly detection methods and can also be substituted for the GK of the Gaussian Mixture Model (GMM) to form an improved generalized fuzzy model [17,18]. The EK density estimation was selected for the moving foreground detection in the improved CAMShift object tracking [19]. However, its application is considerably limited when compared with the Gaussian function because of the imperfection of the theoretical framework. The reason for selecting the EK was its ease of calculation and the discontinuity, which is different from the GK. Owing to discontinuous function, EK is energy concentrated and its modeling results were found to be effective according to our experiments. Our EK-based coding framework is an improvement of the model type for SMoE.

In our previous work, we had developed a rudimentary study on image coding based on EK, which lacked a complete theory and coding scheme design [20]. In this paper, we utilize the theory of mathematical statistics and kernel density estimation to obtain a complete derivation of EK and its related statistics. We also deduce an accurate covariance-matrix expression of three-dimensional (3-D) EK and its marginal distribution, conditional distribution and conditional mean, which are necessary for *Epanechnikov Mixture Regression* (EMR). Furthermore, our EK derivation in the mixture-of-experts provides for other kernels to be applied into regression theory. Under the Bayesian framework, images are modeled by local experts through *Epanechnikov Mixture Models* (EMMs) with global distribution and then the image can be reconstructed by parameters obtained from EMR, which is a stride from modeling theory to image coding.

The novelties of this paper are: (1) deduction of covariance-matrix form of the 3-D EK and its correlated statistics; (2) improving the EM algorithm using MSE optimization; (3) proposing the 3-D EMR and using it in image coding; (4) Adaptive Mode Selection (AMS) combined EMR with GMR.

The remainder of the paper is organized as follows. Section 2 deduces the covariance-matrix form of the three-dimensional Epanechnikov kernel and its correlated statistics. Section 3 presents the algorithm and visualization of EMR. Then, the modeling effect of EMR compared with GMR is discussed in Section 4. In Section 5, the proposed AMS algorithm with coding details is presented. Experimental results are detailed in Section 6. Finally, conclusions and future work are stated in Section 7.

## 2. Correlated theories of 3-D Epanechnikov kernel

### 2.1. Motivation of the kernel selection

The precondition of SMoE is that image pixels are instantiations of a non-linear and non-stationary random process which can be modeled by GMMs. We explore the possibility of image modeling based on mixture models of other kernels. We select Epanechnikov Kernel (EK) because it is discontinuous, which can make its distribution on a central domain and better simulate the functional influence. Moreover, EK can be easily calculated compared to other triple, quartic, and even cos-form functions. Therefore, under the same assumption as SMoE, we break the limitation of the model type and design *Epanechnikov Mixture Models* (EMMs) to accomplish the regression.

### 2.2. Mixture-of-Experts based on Epanechnikov mixture models

In our work, we define variable $x$, $y$, and $z$ as abscissa value, ordinate value, and gray value of each pixel in an image, respectively. As for the formula expression in this section, we define $\boldsymbol{\varphi} = (x, y, z)^T$ and $\boldsymbol{\delta} = (x, y)^T$.

Under the Bayesian framework, we assume that image pixels are modeled as local experts with global distributed 3-D EK, in which the distribution of each EK is global but discontinuous, and thus actually simulates local support. The mixture of Epanechnikov distribution is represented as

$$p(\boldsymbol{\varphi}) = \sum_{j=1}^{K} \alpha_j f_{\boldsymbol{\mu}_j, \boldsymbol{\Sigma}_j}(\boldsymbol{\varphi}), \tag{1}$$

where $\alpha_j$ is the prior corresponding to $\sum_{j=1}^{K} \alpha_j = 1$. $f_{\boldsymbol{\mu}_j, \boldsymbol{\Sigma}_j}(\boldsymbol{\varphi})$ is the 3-D EK whose parameters are $\boldsymbol{\mu}_j$ and $\boldsymbol{\Sigma}_j$. $\boldsymbol{\mu}_j = (\mu_{X_j}, \mu_{Y_j}, \mu_{Z_j})^T$ and $\hat{\boldsymbol{\mu}}_j = (\mu_{X_j}, \mu_{Y_j})^T$ are the mean values of the variables for the $j$th expert. The covariance matrix of the variables for the $j$th expert is $\boldsymbol{\Sigma}_j = \begin{bmatrix} \Sigma_{X_j X_j} & \Sigma_{X_j Y_j} & \Sigma_{X_j Z_j} \\ \Sigma_{Y_j X_j} & \Sigma_{Y_j Y_j} & \Sigma_{Y_j Z_j} \\ \Sigma_{Z_j X_j} & \Sigma_{Z_j Y_j} & \Sigma_{Z_j Z_j} \end{bmatrix}$, within which $\boldsymbol{R}_j = \begin{bmatrix} \Sigma_{X_j X_j} & \Sigma_{X_j Y_j} \\ \Sigma_{Y_j X_j} & \Sigma_{Y_j Y_j} \end{bmatrix}$. Therefore, parameter set of the $j$th mixture model is $\Omega_j = (\alpha_j, \boldsymbol{\mu}_j, \boldsymbol{\Sigma}_j)$.

The regression foundation is the *Mixture-of-Experts* (ME), in which every expert $m_{\boldsymbol{\mu}_j, \boldsymbol{\Sigma}_j}(\boldsymbol{\delta})$ and its gate function $g_{\hat{\boldsymbol{\mu}}_j, \boldsymbol{R}_j, \alpha_j}(\boldsymbol{\delta})$ co-operate with each other [21,22]. $\tilde{m}(\boldsymbol{\delta})$ indicates the expected gray value towards each pixel, from which we reconstruct the decoded image with the parameter sets $\Omega_j = (\alpha_j, \boldsymbol{\mu}_j, \boldsymbol{\Sigma}_j), j = 1, 2, \ldots, K$.

$$\tilde{m}(\boldsymbol{\delta}) = \sum_{j=1}^{K} g_{\hat{\boldsymbol{\mu}}_j, \boldsymbol{R}_j, \alpha_j}(\boldsymbol{\delta}) m_{\boldsymbol{\mu}_j, \boldsymbol{\Sigma}_j}(\boldsymbol{\delta}), \tag{2}$$

Therefore, to use (2) for regression, we have to determine the expression of 3-D EK $f_{\boldsymbol{\mu}_j, \boldsymbol{\Sigma}_j}(\boldsymbol{\varphi})$, gate function $g_{\hat{\boldsymbol{\mu}}_j, \boldsymbol{R}_j, \alpha_j}(\boldsymbol{\delta})$, and conditional mean $m_{\boldsymbol{\mu}_j, \boldsymbol{\Sigma}_j}(\boldsymbol{\delta})$, which are shown in Section 2.2.1, Section 2.2.2, and Section 2.2.3, respectively.

#### 2.2.1. 3-D Epanechnikov kernel

In order to use the model for our compression method, we need to obtain the general function of 3-D EK with respect to the mean vector $\boldsymbol{\mu}_j$ and covariance matrix $\boldsymbol{\Sigma}_j$. We determine the expression of 3-D EK according to the principle that a kernel function $K(\boldsymbol{x})$ must satisfy $\int K(\boldsymbol{x}) d\boldsymbol{x} = 1$ and $K(\boldsymbol{x}) \geq 0$ [23]. For calculation convenience, we begin with a basic ellipsoid form in (3) with axial lengths $a$, $b$, and $c$,

$$\hat{f}(\hat{\boldsymbol{\varphi}}) = \begin{cases} k(1 - \hat{\boldsymbol{\varphi}}^T \boldsymbol{\Lambda}^2 \hat{\boldsymbol{\varphi}}), & \hat{\boldsymbol{\varphi}}^T \boldsymbol{\Lambda}^2 \hat{\boldsymbol{\varphi}} \leq 1 \\ 0, & \text{otherwise}. \end{cases} \tag{3}$$

where $\hat{\boldsymbol{\varphi}} = (\hat{x}, \hat{y}, \hat{z})^T$ represents the variables in this coordinate system. $k$ is the undetermined coefficient for normalization. $\boldsymbol{\Lambda} = \text{diag}(\frac{1}{a}, \frac{1}{b}, \frac{1}{c})$.

The integral of $\hat{f}(\hat{\boldsymbol{\varphi}})$ over its distribution domain is 1, from which we can obtain

$$k \int \int_{\Omega_{\hat{\boldsymbol{\varphi}}}} \int (1 - \hat{\boldsymbol{\varphi}}^T \boldsymbol{\Lambda}^2 \hat{\boldsymbol{\varphi}}) d\Omega_{\hat{\boldsymbol{\varphi}}} = 1, \quad \{\Omega_{\hat{\boldsymbol{\varphi}}} : \hat{\boldsymbol{\varphi}}^T \boldsymbol{\Lambda}^2 \hat{\boldsymbol{\varphi}} \leq 1\}. \tag{4}$$

From Appendix A, it can be observed that $\int \int_{\Omega_{\hat{\boldsymbol{\varphi}}}} \int (1 - \hat{\boldsymbol{\varphi}}^T \boldsymbol{\Lambda}^2 \hat{\boldsymbol{\varphi}}) d\Omega_{\hat{\boldsymbol{\varphi}}} = (8\pi)/(15|\boldsymbol{\Lambda}|)$. As such, we can calculate the undetermined coefficient $k$ as $k = (15|\boldsymbol{\Lambda}|)/(8\pi)$ and the 3-D EK expression with a basic ellipsoid form is

$$\hat{f}(\hat{\boldsymbol{\varphi}}) = \begin{cases} \frac{15|\boldsymbol{\Lambda}|}{8\pi}(1 - \hat{\boldsymbol{\varphi}}^T \boldsymbol{\Lambda}^2 \hat{\boldsymbol{\varphi}}), & \hat{\boldsymbol{\varphi}}^T \boldsymbol{\Lambda}^2 \hat{\boldsymbol{\varphi}} \leq 1 \\ 0, & \text{otherwise}. \end{cases} \tag{5}$$

We define the covariance matrix of $\hat{f}(\hat{\boldsymbol{\varphi}})$ as $\hat{\boldsymbol{\Sigma}}$, whose calculation is detailed in Appendix B. From Appendix B, it is known that $\hat{\boldsymbol{\Sigma}} =$

$\frac{1}{7}(\mathbf{\Lambda}^{-1})^2$, therefore we can get

$$\hat{f}(\hat{\boldsymbol{\varphi}}) = \begin{cases} \frac{15|\mathbf{\Lambda}|}{8\pi}\left(1 - \hat{\boldsymbol{\varphi}}^T \cdot \frac{1}{7}\hat{\boldsymbol{\Sigma}}^{-1} \cdot \hat{\boldsymbol{\varphi}}\right), & \frac{1}{7}\hat{\boldsymbol{\varphi}}^T\hat{\boldsymbol{\Sigma}}^{-1}\hat{\boldsymbol{\varphi}} \leq 1 \\ 0, & \text{otherwise.} \end{cases} \quad (6)$$

Next, we focus on determining the general form of 3-D EK. Because the scale change has been included in $\hat{f}(\hat{\boldsymbol{\varphi}})$, we consider a linear transformation including translation and rotation in (7). The translation distance is $\boldsymbol{\mu}_j = (\mu_{X_j}, \mu_{Y_j}, \mu_{Z_j})^T$ and the orthonormal matrix is $\mathbf{U}$, which satisfies $\mathbf{U}^{-1} = \mathbf{U}^T$ and $\det(\mathbf{U}) = 1$.

$$\hat{\boldsymbol{\varphi}} = \mathbf{U} \cdot (\boldsymbol{\varphi} - \boldsymbol{\mu}_j). \quad (7)$$

Substituting (7) into (5), we obtain the general form of 3-D EK with respect to the linear transformations as follows:

$$\hat{f}(\mathbf{U}(\boldsymbol{\varphi} - \boldsymbol{\mu}_j)) = \begin{cases} \frac{15|\mathbf{\Lambda}|}{8\pi}\left[1 - (\boldsymbol{\varphi} - \boldsymbol{\mu}_j)^T\mathbf{U}^{-1}\mathbf{\Lambda}^2\mathbf{U}(\boldsymbol{\varphi} - \boldsymbol{\mu}_j)\right] & (\boldsymbol{\varphi} - \boldsymbol{\mu}_j)^T\mathbf{U}^{-1}\mathbf{\Lambda}^2\mathbf{U}(\boldsymbol{\varphi} - \boldsymbol{\mu}_j) \leq 1 \\ 0, & \text{otherwise.} \end{cases} \quad (8)$$

The general form of covariance matrix is $\boldsymbol{\Sigma}_j$. Because the general form is acquired by a variable transformation in (7), the covariance matrix is

$$\frac{1}{7}\boldsymbol{\Sigma}_j^{-1} = \mathbf{U}^{-1}\mathbf{\Lambda}^2\mathbf{U}. \quad (9)$$

From (9), we find $|\mathbf{\Lambda}| = 1/(\sqrt{7^3|\boldsymbol{\Sigma}_j|})$ and the general form of three-dimensional Epanechnikov kernel respect to $\boldsymbol{\mu}_j$ and $\boldsymbol{\Sigma}_j$ is

$$f_{\boldsymbol{\mu}_j,\boldsymbol{\Sigma}_j}(\boldsymbol{\varphi}) = \begin{cases} \frac{15}{8\pi\sqrt{7^3|\boldsymbol{\Sigma}_j|}}\left[1 - \frac{1}{7}(\boldsymbol{\varphi} - \boldsymbol{\mu}_j)^T\boldsymbol{\Sigma}_j^{-1}(\boldsymbol{\varphi} - \boldsymbol{\mu}_j)\right] & \frac{1}{7}(\boldsymbol{\varphi} - \boldsymbol{\mu}_j)^T\boldsymbol{\Sigma}_j^{-1}(\boldsymbol{\varphi} - \boldsymbol{\mu}_j) \leq 1 \\ 0, & \text{otherwise.} \end{cases} \quad (10)$$

The Cholesky decomposition of the covariance matrix is also determined:

$$\boldsymbol{\Sigma}_j = \frac{1}{7}\mathbf{U}^{-1}(\mathbf{\Lambda}^{-1})^2\mathbf{U} = \left(\frac{1}{\sqrt{7}}\mathbf{U}^T\mathbf{\Lambda}^{-1}\right)\left(\frac{1}{\sqrt{7}}\mathbf{U}^T\mathbf{\Lambda}^{-1}\right)^T. \quad (11)$$

*2.2.2. Gate function of 3-D EMM*

The gate function $g_{\hat{\boldsymbol{\mu}}_j,\mathbf{R}_j,\alpha_j}(\boldsymbol{\delta})$ is a normalized mixing weight

$$g_{\hat{\boldsymbol{\mu}}_j,\mathbf{R}_j,\alpha_j}(\boldsymbol{\delta}) = \frac{\alpha_j F_{\hat{\boldsymbol{\mu}}_j,\mathbf{R}_j}(\boldsymbol{\delta})}{\sum_{j=1}^{K}\alpha_j F_{\hat{\boldsymbol{\mu}}_j,\mathbf{R}_j}(\boldsymbol{\delta})}, \quad (12)$$

in which $F_{\hat{\boldsymbol{\mu}}_j,\mathbf{R}_j}(\boldsymbol{\delta})$ is the marginal distribution of $f_{\boldsymbol{\mu}_j,\mathbf{R}_j}(\boldsymbol{\varphi})$. Because the distribution of gray level $z$ at each position $\boldsymbol{\varphi}$ is independent of each other, $F_{\hat{\boldsymbol{\mu}}_j,\mathbf{R}_j}(\boldsymbol{\delta})$ can be obtained through

$$F_{\hat{\boldsymbol{\mu}}_j,\mathbf{R}_j}(\boldsymbol{\delta}) = \int_{-\infty}^{+\infty} f_{\boldsymbol{\mu}_j,\boldsymbol{\Sigma}_j}(\boldsymbol{\varphi})dz. \quad (13)$$

To calculate the integration by variable substitution, we define that

$$\boldsymbol{\varphi}' = (x',y',z')^T = \boldsymbol{\varphi} - \boldsymbol{\mu}_j, \quad \boldsymbol{\delta}' = (x',y')^T = \boldsymbol{\delta} - \hat{\boldsymbol{\mu}}_j. \quad (14)$$

Then, the integral turns out to be

$$F_{\mathbf{0},\mathbf{R}_j}(\boldsymbol{\delta}') = \int_{-\infty}^{+\infty} f_{\mathbf{0},\boldsymbol{\Sigma}_j}(\boldsymbol{\varphi}')dz' = \int_{z_1'}^{z_2'} \frac{15}{8\pi\sqrt{7^3|\boldsymbol{\Sigma}_j|}}\left(1 - \frac{1}{7}\boldsymbol{\varphi}'^T\boldsymbol{\Sigma}_j^{-1}\boldsymbol{\varphi}'\right)dz', \quad (15)$$

in which $z_1'$ and $z_2'$ can be solved from the two endpoints of $\frac{1}{7}\boldsymbol{\varphi}'^T\boldsymbol{\Sigma}_j^{-1}\boldsymbol{\varphi}' = 1$.

To simplify the calculation, we represent the inverse of the covariance matrix as

$$\boldsymbol{\Sigma}_j^{-1} = \begin{bmatrix} \Sigma_{X_jX_j} & \Sigma_{X_jY_j} & \Sigma_{X_jZ_j} \\ \Sigma_{Y_jX_j} & \Sigma_{Y_jY_j} & \Sigma_{Y_jZ_j} \\ \Sigma_{Z_jX_j} & \Sigma_{Z_jY_j} & \Sigma_{Z_jZ_j} \end{bmatrix}^{-1} = \begin{bmatrix} \sigma_a & \sigma_b & \sigma_c \\ \sigma_b & \sigma_d & \sigma_e \\ \sigma_c & \sigma_e & \sigma_f \end{bmatrix}. \quad (16)$$

Therefore, we obtain

$$\frac{1}{7}(x',y',z')\boldsymbol{\Sigma}_j^{-1}(x',y',z')^T = 1 \quad (17)$$

$$\sigma_f z'^2 + 2(\sigma_c x' + \sigma_e y')z' + (\sigma_a x'^2 + \sigma_d y'^2 + 2\sigma_b x'y' - 7) = 0. \quad (18)$$

Hence, we solve the equation (18) and obtain the solutions as

$$z_1' = \frac{-(\sigma_c x' + \sigma_e y') - \sqrt{t}}{\sigma_f}, \quad z_2' = \frac{-(\sigma_c x' + \sigma_e y') + \sqrt{t}}{\sigma_f} \quad (19)$$

in which

$$t = (\sigma_c x' + \sigma_e y')^2 - (\sigma_a x'^2 + \sigma_d y'^2 + 2\sigma_b x'y' - 7)\sigma_f. \quad (20)$$

Finally, when $\frac{1}{7}\boldsymbol{\varphi}'^T\boldsymbol{\Sigma}_j^{-1}\boldsymbol{\varphi}' \leq 1$, the integral in (15) becomes

$$\begin{aligned} F_{\mathbf{0},\mathbf{R}_j}(\boldsymbol{\delta}') &= \frac{15}{8\pi\sqrt{7^3|\boldsymbol{\Sigma}_j|}}\int_{z_1'}^{z_2'}\left(1 - \frac{1}{7}\boldsymbol{\varphi}'^T\boldsymbol{\Sigma}_j^{-1}\boldsymbol{\varphi}'\right)dz' = \frac{15}{8\pi\sqrt{7^3|\boldsymbol{\Sigma}_j|}} \cdot \frac{4t^{\frac{3}{2}}}{21\sigma_f^2} \\ &= \frac{15}{8\pi\sqrt{7^3|\boldsymbol{\Sigma}_j|}} \cdot \frac{4|\boldsymbol{\Sigma}_j|^{\frac{1}{2}}\left[7|\mathbf{R}_j| - (x',y')|\mathbf{R}_j|\mathbf{R}_j^{-1}(x',y')^T\right]^{\frac{3}{2}}}{21|\mathbf{R}_j|^2} \\ &= \frac{5}{14\pi\sqrt{|\mathbf{R}_j|}} \cdot \left(1 - \frac{1}{7}\boldsymbol{\delta}'^T\mathbf{R}_j^{-1}\boldsymbol{\delta}'\right)^{\frac{3}{2}}. \end{aligned} \quad (21)$$

Therefore, the general form of the marginal distribution is

$$F_{\hat{\boldsymbol{\mu}}_j,\mathbf{R}_j}(\boldsymbol{\delta}) = \begin{cases} \frac{5}{14\pi\sqrt{|\mathbf{R}_j|}}\left[1 - \frac{1}{7}(\boldsymbol{\delta} - \hat{\boldsymbol{\mu}}_j)^T\mathbf{R}_j^{-1}(\boldsymbol{\delta} - \hat{\boldsymbol{\mu}}_j)\right]^{\frac{3}{2}}, & \frac{1}{7}(\boldsymbol{\delta} - \hat{\boldsymbol{\mu}}_j)^T\mathbf{R}_j^{-1}(\boldsymbol{\delta} - \hat{\boldsymbol{\mu}}_j) \leq 1 \\ 0, & \text{otherwise.} \end{cases} \quad (22)$$

*2.2.3. Conditional mean*

Because the distribution of $z$ is independent at each location $\boldsymbol{\delta}$, we conclude the conditional distribution of $z|\boldsymbol{\delta}$ as $e_{\boldsymbol{\mu}_j,\boldsymbol{\Sigma}_j}(z|\boldsymbol{\delta})$:

$$e_{\boldsymbol{\mu}_j,\boldsymbol{\Sigma}_j}(z|\boldsymbol{\delta}) = \frac{f_{\boldsymbol{\mu}_j,\boldsymbol{\Sigma}_j}(\boldsymbol{\varphi})}{F_{\hat{\boldsymbol{\mu}}_j,\mathbf{R}_j}(\boldsymbol{\delta})}$$

$$= \begin{cases} \frac{3}{4}\sqrt{\frac{|\mathbf{R}_j|}{|\boldsymbol{\Sigma}_j|}}\frac{1 - \frac{1}{7}(\boldsymbol{\varphi} - \boldsymbol{\mu}_j)^T\boldsymbol{\Sigma}_j^{-1}(\boldsymbol{\varphi} - \boldsymbol{\mu}_j)}{\left[1 - \frac{1}{7}(\boldsymbol{\delta} - \hat{\boldsymbol{\mu}}_j)^T\mathbf{R}_j^{-1}(\boldsymbol{\delta} - \hat{\boldsymbol{\mu}}_j)\right]^{\frac{3}{2}}} & 1 - \frac{1}{7}(\boldsymbol{\varphi} - \boldsymbol{\mu}_j)^T \\ & \times \boldsymbol{\Sigma}_j^{-1}(\boldsymbol{\varphi} - \boldsymbol{\mu}_j) \leq 1 \\ 0, & \text{otherwise.} \end{cases} \quad (23)$$

Combined with (14), the conditional mean of $z$ in the condition of $\boldsymbol{\delta}$ is

$$m_{\mathbf{0},\boldsymbol{\Sigma}_j}(\boldsymbol{\delta}') = \int_{-\infty}^{+\infty} z e_{\boldsymbol{\mu}_j,\boldsymbol{\Sigma}_j}(z|\boldsymbol{\delta})dz = \int_{-\infty}^{+\infty}(z' + \mu_{Z_j})e_{\mathbf{0},\boldsymbol{\Sigma}_j}(z'|\boldsymbol{\delta}')dz' \quad (24)$$

We use (19) to obtain the following integral:

$$\begin{aligned} m_{\mathbf{0},\boldsymbol{\Sigma}_j}(\boldsymbol{\delta}') &= \mu_{Z_j} + \int_{z_1'}^{z_2'} z' e_{\mathbf{0},\boldsymbol{\Sigma}_j}(z'|\boldsymbol{\delta}')dz' = \mu_{Z_j} + \int_{z_1'}^{z_2'} z' \frac{f_{\mathbf{0},\boldsymbol{\Sigma}_j}(\boldsymbol{\varphi}')}{F_{\mathbf{0},\mathbf{R}_j}(\boldsymbol{\delta}')}dz' \\ &= \mu_{Z_j} + \int_{z_1'}^{z_2'} z' \frac{\frac{15}{8\pi\sqrt{7^3|\boldsymbol{\Sigma}_j|}}\left(1 - \frac{1}{7}\boldsymbol{\varphi}'^T\boldsymbol{\Sigma}_j^{-1}\boldsymbol{\varphi}'\right)}{\frac{15}{8\pi\sqrt{7^3|\boldsymbol{\Sigma}_j|}} \cdot \frac{4t^{\frac{3}{2}}}{21\sigma_f^2}} dz' \\ &= \mu_{Z_j} + \frac{21\sigma_f^2}{4t^{\frac{3}{2}}} \cdot \int_{z_1'}^{z_2'} z'\left(1 - \frac{1}{7}\boldsymbol{\varphi}'^T\boldsymbol{\Sigma}^{-1}\boldsymbol{\varphi}'\right)dz' = \mu_{Z_j} - \frac{\sigma_c x' + \sigma_e y'}{\sigma_f}. \end{aligned} \quad (25)$$

According to (16), the result of the conditional mean can be formulated as

$$\begin{aligned} m_{\boldsymbol{\mu}_j,\boldsymbol{\Sigma}_j}(\boldsymbol{\delta}) &= \mu_{Z_j} - \frac{(\sigma_c,\sigma_e)(\boldsymbol{\delta} - \boldsymbol{\mu}_j)}{\sigma_f} = \mu_{Z_j} \\ &\quad + \left(\Sigma_{Z_jX_j},\Sigma_{Z_jY_j}\right)\mathbf{R}_j^{-1}(\boldsymbol{\delta} - \hat{\boldsymbol{\mu}}_j). \end{aligned} \quad (26)$$

## 3. 3-D Epanechnikov mixture regression

### 3.1. Representation and optimization

We design the EMR in Algorithm 1 to estimate the parame-

---

**Algorithm 1** Epanechnikov Mixture Regression (EMR).

---
1: Initialize matrix $\mathbf{data}_{N \times 3}$ using *Kmeans algorithm* into $K$ clusters, whose parameters are $\mathbf{\Omega}_1 = (\boldsymbol{\mu}_{j1}, \boldsymbol{\Sigma}_{j1}, \alpha_{j1})$, $j = 1, 2, \ldots, K$.
2: Substitute $\mathbf{\Omega}_1$ into (2) and obtain the initial modeled block $\hat{Z}_1$. Then, we can obtain MSE(1) $= \frac{1}{N} \sum_{i=1}^{N} (z_{\hat{Z}_{1i}} - z_{\hat{Z}_{0i}})^2$.
3: **for** $t = 2 : 8$
4:    Update $Qt_{ij}$, $i = 1, 2, \ldots, N$, $j = 1, 2, \ldots, K$ using (27).
5:    **for** $j = 1 : K$
6:       Update $\boldsymbol{\mu}_{jt} = \frac{\sum_{i=1}^{N} Qt_{ij} \boldsymbol{\varphi}_i}{\sum_{i=1}^{N} Qt_{ij}}$, $\mathbf{\Sigma}_{jt} = \frac{\sum_{i=1}^{N} Qt_{ij}(\boldsymbol{\varphi}_i - \boldsymbol{\mu}_j)^T (\boldsymbol{\varphi}_i - \boldsymbol{\mu}_j)}{\sum_{i=1}^{N} Qt_{ij}}$, $\alpha_{jt} = \frac{\sum_{i=1}^{N} Qt_{ij}}{N}$.
7:    **end**
8:    Substituting $\mathbf{\Omega}_t = (\boldsymbol{\mu}_{jt}, \mathbf{\Sigma}_{jt}, \alpha_{jt})$, $j = 1, 2, \ldots, K$ into (2), we can obtain MSE(t) $= \frac{1}{N} \sum_{i=1}^{N} (z_{\hat{Z}_{ti}} - z_{\hat{Z}_{0i}})^2$.
9: **end**
10: $t_0 =$ find (MSE == min(MSE))
11: **Output** $\mathbf{\Omega}_{t_0}$

---

ters of EMM for optimizing the gray value in each image pixel. We model the Y, U, and V channels of a color image separately. As for a certain channel, it is assumed that the total number of channel pixels is $N$ and the $i$th pixel is denoted as $(x_i, y_i, z_i)$, in which $x_i$, $y_i$, and $z_i$ represent the X-coordinate, Y-coordinate, and gray value of the pixel, respectively. We represent each channel this way to prepare for modeling the next step. Therefore, we can represent the data of each channel as an $N \times 3$ matrix as $\mathbf{data}_{N \times 3}$ and then we use *Kmeans++ Clustering* [24] to initialize $\mathbf{data}_{N \times 3}$ into $K$ models whose parameters are $\mathbf{\Omega}_1 = (\boldsymbol{\mu}_{j1}, \boldsymbol{\Sigma}_{j1}, \alpha_{j1})$, $j = 1, 2, \ldots, K$.

We apply the M-step of the *Expectation-Maximization* (EM) algorithm [25] of GMM in Step 6 of Algorithm 1 to iterate for the optimal EMM parameters. The core of E-step is the posterior probability

$$Qt_{ij} = P(j|\boldsymbol{\varphi}_i) = \frac{\alpha_j f_{\boldsymbol{\mu}_j, \boldsymbol{\Sigma}_j}(\boldsymbol{\varphi}_i)}{\sum_{j=1}^{K} \alpha_j f_{\boldsymbol{\mu}_j, \boldsymbol{\Sigma}_j}(\boldsymbol{\varphi}_i)}. \tag{27}$$

Unalike the original EM algorithm, we do not use the log-likelihood for regression. In this study, we measure the *Mean Square Error* (MSE) between the recovery $\hat{Z}_t$ in (2) and the original image block $\hat{Z}_0$ for each iteration of parameters and select the parameter set with the minimum MSE. $z_{\hat{Z}_{0i}}$ in Algorithm 1 indicates the gray value of the $i$th pixel in block $\hat{Z}_0$, and $z_{\hat{Z}_{ti}}$ indicates the gray value of the $i$th pixel of the $t$th iteration block $\hat{Z}_t$.

The two EM optimization algorithms are shown in Fig. 1, and it shows that the MSE based algorithm has better performance and consumes less time compared to the log-like-based optimization. This is because our MSE based algorithm directly selects the reconstructed image according to the error to the original one and there is also no time loss for calculating the log-like probability.

### 3.2. Theoretical visualization of regression

We select a $32 \times 32$ block of *Lena* modeled by nine kernels using EMR and GMR for visualizing the kernel comparison examples in Fig. 2. To ensure that the experiments are fair, we implement the mixture regression of GMM identical to Algorithm 1 only by substituting (27) with

$$Qt_{ij} = p(j|\boldsymbol{\varphi}_i) = \frac{\alpha_j \cdot \frac{1}{\sqrt{(2\pi)^3 |\boldsymbol{\Sigma}_j|}} \exp\left[-\frac{1}{2} (\boldsymbol{\varphi}_i - \boldsymbol{\mu}_j)^T \boldsymbol{\Sigma}_j^{-1} (\boldsymbol{\varphi}_i - \boldsymbol{\mu}_j)\right]}{\sum_{j=1}^{K} \alpha_j \cdot \frac{1}{\sqrt{(2\pi)^3 |\boldsymbol{\Sigma}_j|}} \exp\left[-\frac{1}{2} (\boldsymbol{\varphi}_i - \boldsymbol{\mu}_j)^T \boldsymbol{\Sigma}_j^{-1} (\boldsymbol{\varphi}_i - \boldsymbol{\mu}_j)\right]}. \tag{28}$$

Fig. 2(d) outlines the continuous probability distribution in (1), in which the nine ellipsoids represent the nine models with Epanechnikov distribution. Compared to GMM in Fig. 2(g), EMM has a tighter correlated distribution.

The concentrated-distribution ellipsoids of EMM can enhance the correlation of regression. According to (2), the two essential statistics for the image reconstruction are: the conditional mean $m_{\boldsymbol{\mu}_j, \boldsymbol{\Sigma}_j}(\boldsymbol{\delta})$ and gate function $g_{\hat{\boldsymbol{\mu}}_j, \boldsymbol{R}_j, \alpha_j}(\boldsymbol{\delta})$ of each Epanechnikov expert, which can be respectively visualized through (26) and (12). Conditional means $m_{\boldsymbol{\mu}_j, \boldsymbol{\Sigma}_j}(\boldsymbol{\delta})$, $j = 1, 2, \ldots, 9$ of EMR are shown as nine linear mean planes in Fig. 2(e), whose slope is steeper than that of the GMR in Fig. 2(h). The reason is that the EK function is locally simulated, whereas GK has a flat long tail of distribution. Therefore, the covariance matrix of EK may be more extreme and then the gray value has a shaper change according to (26). The top view of $g_{\hat{\boldsymbol{\mu}}_j, \boldsymbol{R}_j, \alpha_j}(\boldsymbol{\delta})$, $j = 1, 2, \ldots, 9$ is shown in Fig. 2(f), in which the distinct borders of each gate function of EMR can be seen, whereas the gate function of GMR shown in Fig. 2(i) is smooth. This reflects that the discontinuity of EK-based function simulates the local support of gate function, which makes the reconstructed image of EMR clearer than that of GMR.

## 4. EMR Modeling representation

In this section, we analyze the modeling effects of the two kernels by conducting experiments. The EMR is in Algorithm 1 and the optimization framework of GMR has been the same with EMR as mentioned in Section 3.2. In this part of the experiments, we model an image block using EMR and GMR under the blocksizes of $16 \times 16$, $32 \times 32$, and $64 \times 64$ using different quantities of models. In addition, we estimate the bit consumption of a single kernel towards different blocksizes, which is shown in Table 1. The statistics, which are the same for EMR and GMR, are obtained from hundreds of image pieces with accurate quantization and the reconstructed block is nearly the same as the modeling result. The modeling results in Fig. 3 include bit-SSIM curves and comparison examples.

Examples for modeling results using eight $256 \times 256$ image blocks are shown in Fig. 3. The ranges of SSIM are slightly different for each image result because some images contain much high frequency contents, whereas others are flat, and the different numbers of models is another reason. From Fig. 3, we can see that EMR shows better performance over GMR at $16 \times 16$ blocks based modeling, and EMR16 is always the best of all. If the number of experimental models is increased, the improvements of image quality will be saturated, which is shown in Fig. 3 (d). These results confirm the theoretical conclusion described in Section 3.2: the discontinuity of EMR simulates the local effect of modeling; therefore, the smaller blocks have better performance, whereas the modeling of larger blocks cannot be well correlated.

As for the running speed of both kernels, we present the sum of modeling and reconstruction times of *Butterfly* as an example. If the number of models for each $16 \times 16$ block is 11, the total time consumed by the entire image is 9.87 s and 10.89 s for EMR and GMR, respectively. For $32 \times 32$ based modeling, if the number of models is 38 for each block, the total run time is 21.05 s and 20.44 s for EMR and GMR, respectively. The modeling of the entire image based on a $64 \times 64$ block requires 170.00 s and 181.17 s by EMR and GMR, respectively, when we use 144 models in each

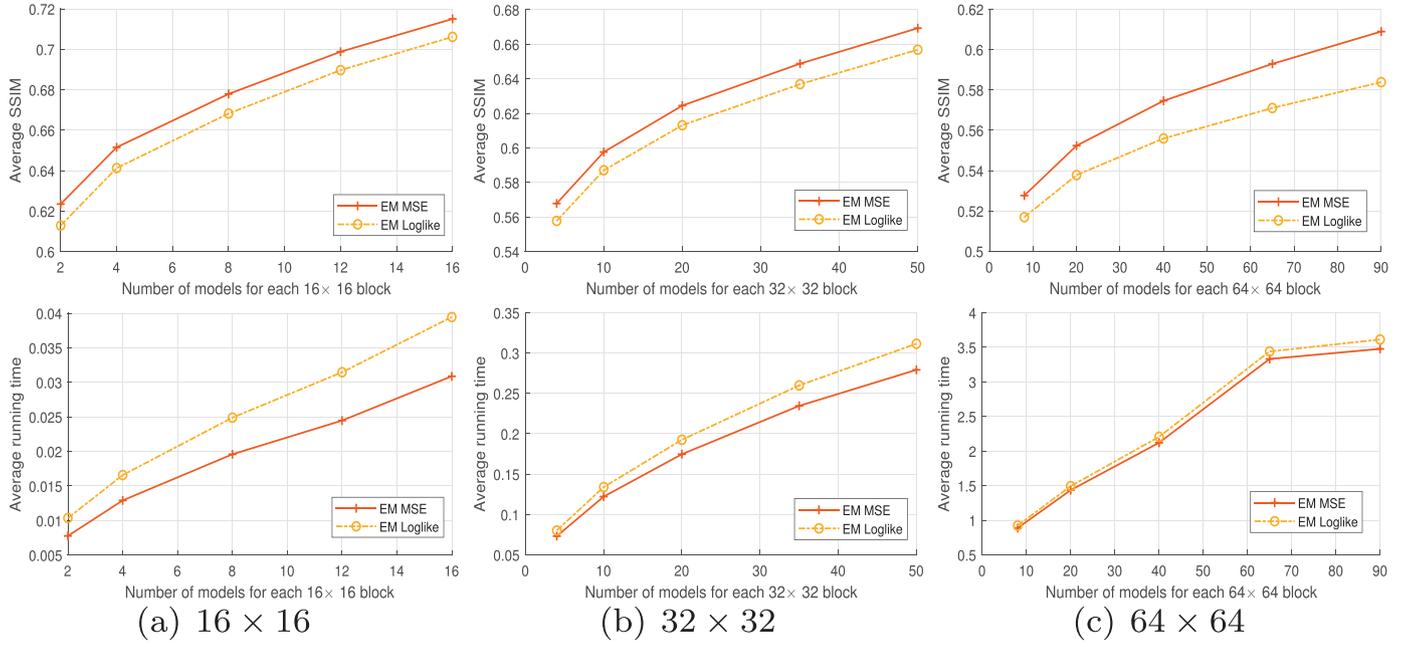

**Fig. 1.** Average SSIM and time consuming comparing among three EM optimization algorithms for a certain block of size 16 × 16, 32 × 32, and 64 × 64. Our MSE-based EM algorithm is the optimal one.

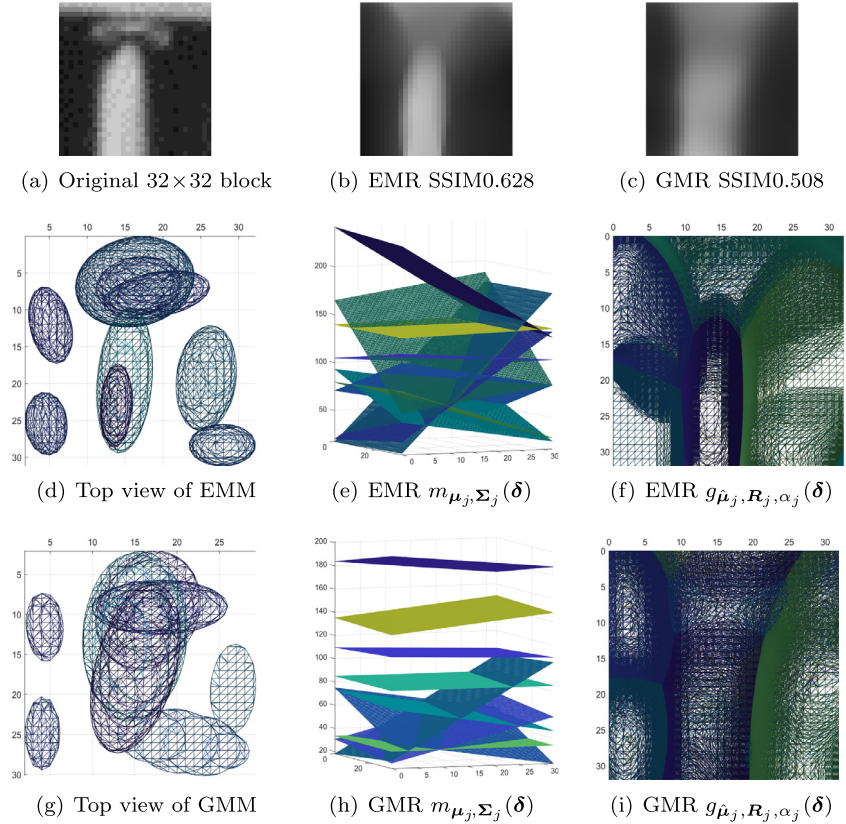

**Fig. 2.** Experimental results modeled by EMR and GMR together with the regressing visualization.

**Table 1**
Bits Consumption Towards Three Blocksizes of EMM/GMM.

| Blocksize | $\mu_{X_j}$ | $\mu_{Y_j}$ | $\mu_{Z_j}$ | $\Sigma_{X_j X_j}$ | $\Sigma_{X_j Y_j}$ | $\Sigma_{X_j Z_j}$ | $\Sigma_{Y_j Y_j}$ | $\Sigma_{Y_j Z_j}$ | $\alpha_j$ | bits/kernel |
|---|---|---|---|---|---|---|---|---|---|---|
| 16 × 16 | 4 | 4 | 8 | 6 | 7 | 10 | 6 | 10 | 7 | 62 |
| 32 × 32 | 5 | 5 | 8 | 8 | 9 | 11 | 8 | 11 | 7 | 72 |
| 64 × 64 | 6 | 6 | 8 | 10 | 10 | 12 | 10 | 12 | 7 | 81 |

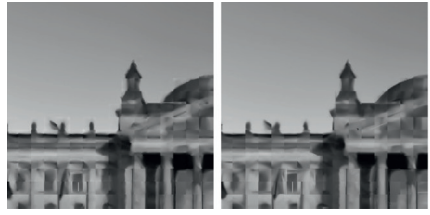

SSIM$_{EMR}$=0.798  SSIM$_{GMR}$=0.794

Blocksize = 16, kernels = 8

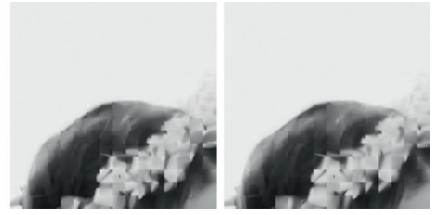

SSIM$_{EMR}$=0.797  SSIM$_{GMR}$=0.791

Blocksize = 16, kernels = 5

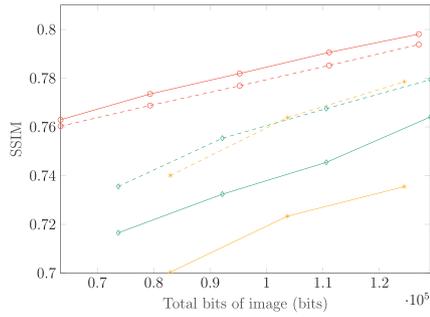

(a) *Architecture*. The number of models for each $16 \times 16$ is 4,5,6,7,8; each $32 \times 32$ is 16,20,24,28; each $64 \times 64$ is 64,80,96.

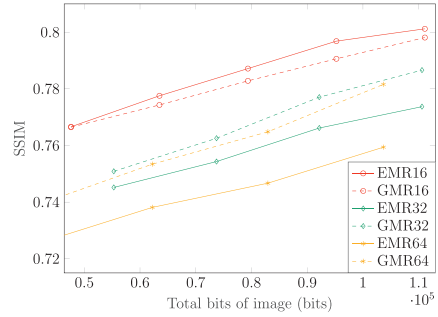

(b) *Garland*. The number of models for each 16×16 is 3,4,5,6,7; each 32×32 is 12,16,20,24; each 64×64 is 32,48,64,80.

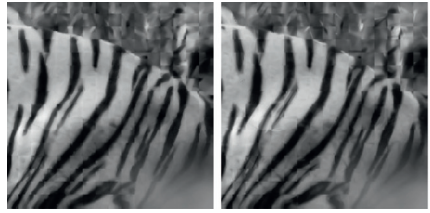

SSIM$_{EMR}$=0.746  SSIM$_{GMR}$=0.731

Blocksize = 16, kernels = 13

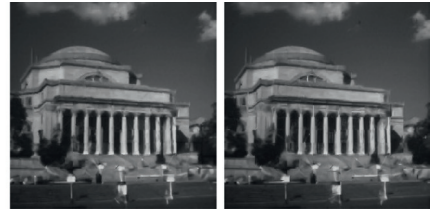

SSIM$_{EMR}$=0.897  SSIM$_{GMR}$=0.884

Blocksize = 16, kernels = 25

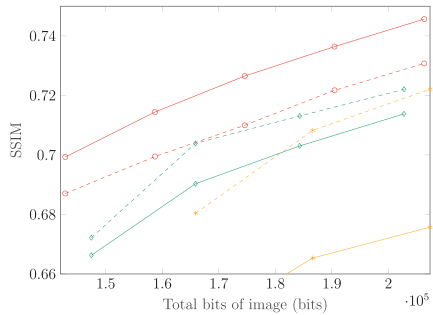

(c) *Tiger*. The number of models for each $16 \times 16$ is 9,10,11,12,13; each $32 \times 32$ is 32,36,40,44; each $64 \times 64$ is 128,144,160.

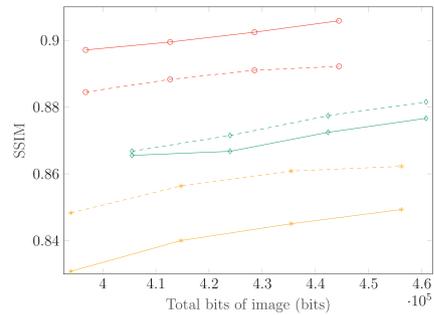

(d) *Columbia*. The number of models for each $16 \times 16$ is 25,26,27,28; each $32 \times 32$ is 88,92,96,100; each $64 \times 64$ is 304,320,336,352.

**Fig. 3.** Modeling results. The numbers of models used for bit-SSIM curves are marked at each sub-caption.

block. Therefore, the speed is similar for both kernels and the smaller block pattern is more efficient.

## 5. Coding framework

As shown in our schematic diagram in Fig. 4, a color image is encoded in Y, U, and V channels separately, within which the U and V channels are downsampled once in both directions into U$_D$ and V$_D$. $\lambda$, the same for Y and U/V channel, is the Lagrange parameter in (29) which we use to control the adaptive modeling. To realize a smooth transition from adjacent blocks, we implement the Adaptive Mode Selection (AMS) algorithm with overlapping blocks referenced in [9], and we set that the width as $OL_Y$ for the Y channel, and $OL_{UV}$ for the U/V channel. As for a certain channel at

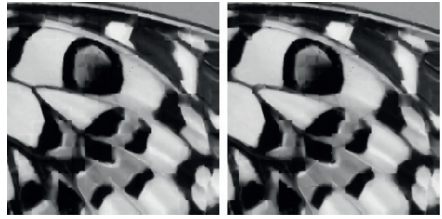

$SSIM_{EMR} = 0.770$    $SSIM_{GMR} = 0.763$

Blocksize = 16, kernels = 11

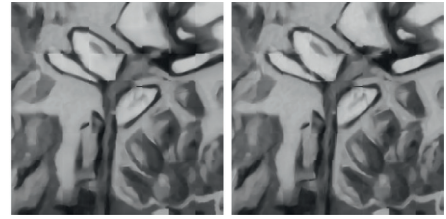

$SSIM_{EMR} = 0.772$    $SSIM_{GMR} = 0.792$

Blocksize = 32, kernels = 48

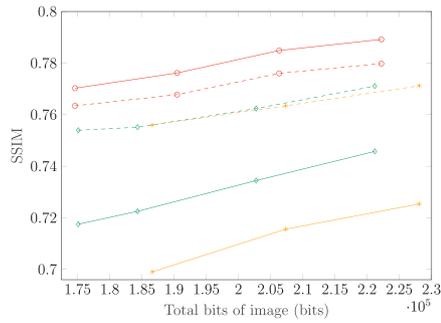

(e) *Butterfly*. The number of models for each $16 \times 16$ is 11,12,13,14; each $32 \times 32$ is 38,40,44,48; each $64 \times 64$ is 144,160,176.

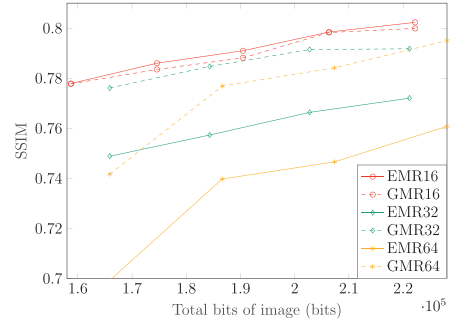

(f) *Flowers*. The number of models for each $16 \times 16$ is 10,11,12,13,14; each $32 \times 32$ is 36,40,44,48; each $64 \times 64$ is 128,144,160,176.

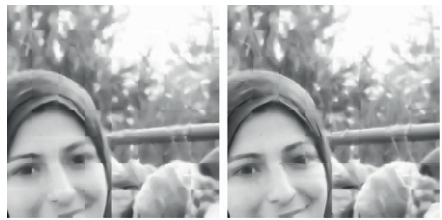

$SSIM_{EMR} = 0.770$    $SSIM_{GMR} = 0.796$

Blocksize = 32, kernels = 36

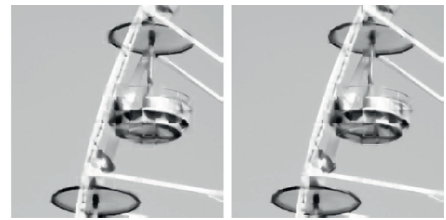

$SSIM_{EMR} = 0.908$    $SSIM_{GMR} = 0.904$

Blocksize = 16, kernels = 12

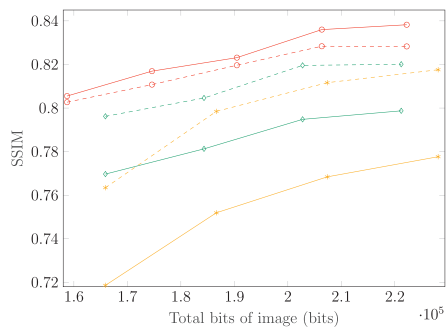

(g) *Girl*. The number of models for each $16 \times 16$ is 10,11,12,13,14; each $32 \times 32$ is 36,40,44,48; each $64 \times 64$ is 128,144,160,176.

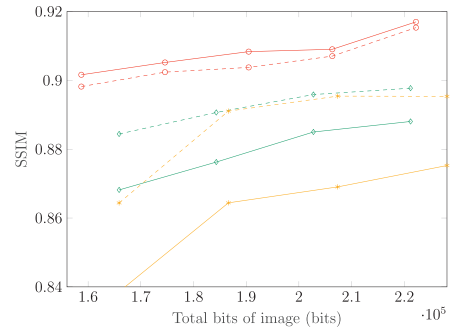

(h) *Wheel*. The number of models for each $16 \times 16$ is 10,11,12,13,14; each $32 \times 32$ is 36,40,44,48; each $64 \times 64$ is 128,144,160,176.

**Fig. 3.** Continued

a certain $\lambda$, different sizes of blocks have the same overlapping width.

In the decoder, the reconstructed $U_{Dr}$ and $V_{Dr}$ are de-blocked and upsampled into $U_r$ and $V_r$ just by replicating the adjacent pixel in a row or column.

### 5.1. Adaptive mode selection

We use the EMR and GMR mentioned in Section 4 to design the AMS algorithm and to determine an optimal mode of coding, whose goal is to achieve the optimal coding mode satisfying the balance between image distortion and bitrates. As for the coding mode in our framework, each 64 × 64 block can be divided into 32 × 32 and 16 × 16 units encoded by either GMR or EMR with an adequate number of models.

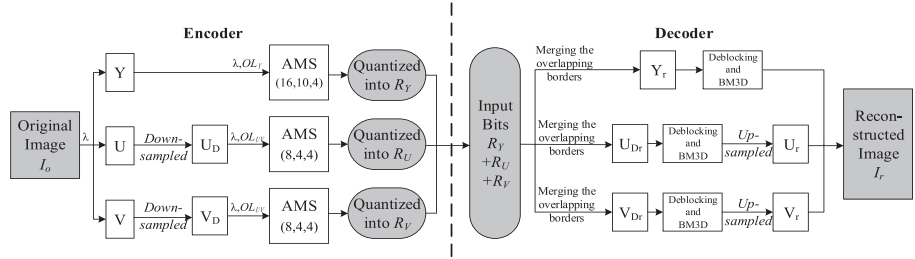

**Fig. 4.** A schematic diagram of our coding framework.

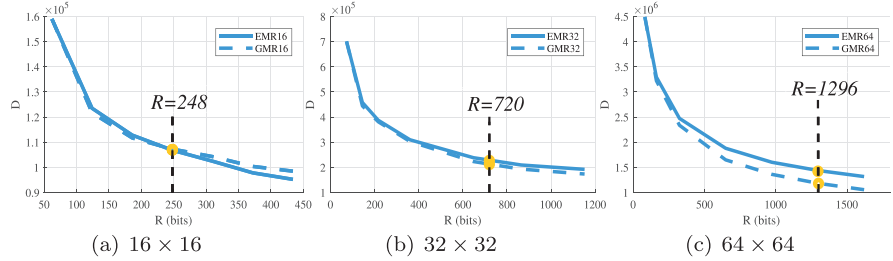

**Fig. 5.** R-D curves of EMR and GMR based coding respectively towards 16 × 16, 32 × 32 and 64 × 64 blocks. The black dotted line is the probable critical point of R.

Using $J$ in (29), our AMS algorithm selects the mode corresponding to the minimum $J$. The specific task is to minimize the distortion $D$ subject to a selected bit rate $R$ for an image block of a particular size. The constrained problem turns out to be an unconstrained one and we aimed at determining a reasonable $\lambda$ to minimize the value of $J$ expressed in (29) by using the Lagrange multiplier method.

$$J = D + \lambda R, \qquad (29)$$

in which $\lambda = -\partial D/\partial R$ is the negative slope of the R-D curve, which represents the Lagrange multiplier. $D$ is the squared error, and $R$ is the total number of bits.

The bits consuming of accurate modeling for each kernel is 62, 72, and 81 towards 16×16, 32×32, and 64×64 blocks, respectively, as shown in Table 1. We can then use thousands of image pieces to estimate relatively accurate R-D curves as shown in Fig. 5, from which we estimate the critical bitrate. The critical bitrate is a bitrate value that can satisfy the balance we want to achieve between image distortion and bit consumption, which is marked as a black dotted line in Fig. 5. As for 16 × 16, 32 × 32, and 64 × 64 blocks, we first set the approximate critical bitrate to be 248, 720, and 1296 bits, respectively, equivalent to the maximum number of models iterations to be $248/62 = 4$, $720/72 = 10$, and $1296/81 = 16$. Therefore, we set the upper limit of number of models to ($n64, n32, n16$) of Y channels as 16, 10, and 4. Moreover, the upper limit ($n64, n32, n16$) of U and V channels is set as 8, 4, and 4.

The pseudo-code of AMS is given in Algorithm 2, in which **J** represents the J-values of the modeled block regressed by Epanechnikov or Gaussian kernel with different numbers of models, that is, $J_{E_3}$ in Step 2 denotes J-value of 64 × 64 block modeled by Epanechnikov kernel using three models. Notably, the $D$ in (29) is calculated without the added overlapping parts. Our algorithm iterates over all the cases through nested loops and the final result is produced by filtering the optimal result of each loop according to the minimum value of $J$. The mode includes block size, kernel type, and the number of experts drawn through command *find* from matrix **J**. From the conclusion detailed in Section 4, we just model the 16×16 and 64×64 blocks using EMR and GMR, re-

---

**Algorithm 2** Adaptive Mode Selection: AMS ($n64, n32, n16$).

1: **Input:** 64 × 64 image block $A$, a target $\lambda$
2: $\mathbf{J} = [J_{E_1} J_{E_2} J_{E_3} \cdots J_{E_{n_{64}}}]$ and $J_{64} = \min\mathbf{J}$, $\text{mode}_{64} = \text{find}(\mathbf{J} == J_{64})$
3: **for** $A_i$, $i = 1 : 4$ (Divide $A$ into 4 32 × 32 blocks ($A_1$ to $A_4$))
4: $\mathbf{J} = \begin{bmatrix} J_{E_1} J_{E_2} J_{E_3} \cdots J_{E_{n_{32}}} \\ J_{G_1} J_{G_2} J_{G_3} \cdots J_{G_{n_{32}}} \end{bmatrix}$ and $J_{32_i} = \min\mathbf{J}$, $\text{mode}_{32_i} = \text{find}(\mathbf{J} == J_{32_i})$
5:   **for** $A_{ij}$, $j = 1 : 4$ (Divide $A_i$ into 4 16 × 16 pieces ($A_{i1}$ to $A_{i4}$))
6:     $\mathbf{J} = [J_{E_1} J_{E_2} J_{E_3} \cdots J_{E_{n_{16}}}]$ and $J_{16_{ij}} = \min\mathbf{J}$, $\text{mode}_{16_{ij}} = \text{find}(\mathbf{J} == J_{16_{ij}})$
7:   **end**
8:   Compare $\sum_{j=1}^{4} J_{16_{ij}}$ with $J_{32_i}$, which are respectively corresponding to $\begin{bmatrix} \text{mode}_{16_{i1}} & \text{mode}_{16_{i2}} \\ \text{mode}_{16_{i3}} & \text{mode}_{16_{i4}} \end{bmatrix}$ and $\text{mode}_{32_i}$. The smaller value is set as $J'_{32_i}$ and the corresponding modes are recoded as $\text{mode}'_{32_i}$.
9: **end**
10: Like Step 8, we pick up the mode between $\begin{bmatrix} \text{mode}_{32_1} & \text{mode}_{32_2} \\ \text{mode}_{32_3} & \text{mode}_{32_4} \end{bmatrix}$ and $\text{mode}_{64}$ by choosing the smaller one between $\sum_{i=1}^{4} J'_{32_i}$ and $J_{64}$.
11: **Output:** Parameters of the optimal mode of modeling combination.

---

spectively. However, we make a selection between EMR and GMR at the 32×32 based condition because of the comparable effect. Examples of mode selection are shown in Fig. 6, and the model distribution under different conditions can be observed.

### 5.2. Coding of modeling parameters

A certain image has a header that contains marks of the minimum value and span of all the parameters except $\eta_j$, the angle of eigen-decomposition in (30). $\eta_j$ does not require a header mark because its range is certainly $[-90°, 90°]$. Therefore, for a certain Y, U, or V channel, a header has $3 \times 3 \times 7 \times 2 \times 8 = 1008$ bits (3 channels, 3 blocksizes, 7 parameters with 2 marks and 8 bits each).

As for the encoding within each block, there is a *Flag* which determines the blocksize ($B$), number of models ($NM$), and the kernel type ($EG$). There is another portion of *Parameters* encoded as $n_P$ according to different blocksizes. The bits for flag and parameters of

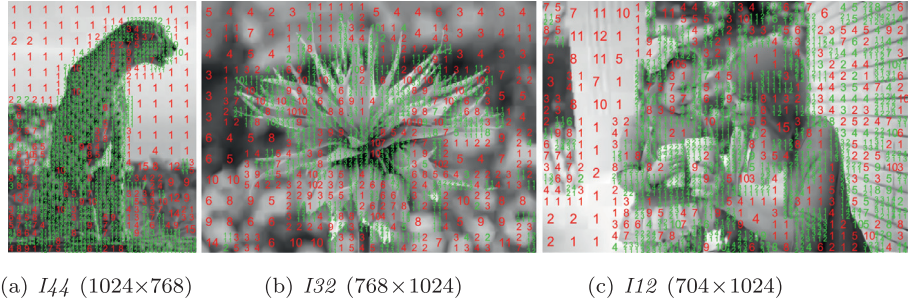

(a) $I44$ (1024×768)  (b) $I32$ (768×1024)  (c) $I12$ (704×1024)

**Fig. 6.** Optimal mode selection examples of our AMS towards Y-channel ($\lambda = 100$). The green numbers represent the number of models of EMR and the red ones represent GMR. (For interpretation of the references to colour in this figure legend, the reader is referred to the web version of this article.)

**Table 2**
Bits Allocation of Parameters of a Certain Kernel for Different Blocksizes: $B$ is encoded as $11$ ($16 \times 16$), $10$ ($32 \times 32$) and $0$ ($64 \times 64$). $n_{NM}$ of different blocksizes depends on the upper limit of the number of models ($n_{16}$, $n_{32}$, and $n_{64}$). $EG$ is encoded as $1$ for EK as well as $0$ for GK.

| Condition | | Kernel Bits Allocation | | | | | | | | | |
|---|---|---|---|---|---|---|---|---|---|---|---|
| | | Flag $n_F$ (bits) | | | Parameters $n_P$ (bits) | | | | | | |
| NM | B | $n_B$ | $n_{NM}$ | $n_{EG}$ | $n_{\mu_{X_j}}$ | $n_{\mu_{Y_j}}$ | $n_{\mu_{Z_j}}$ | $n_{\eta_j}$ | $n_{e_{1_j}}$ | $n_{e_{2_j}}$ | $n_{\Sigma_{X_j Z_j}}$ | $n_{\Sigma_{Y_j Z_j}}$ |
| Y >1 | 16×16 | 2 | 2 | - | 3 | 3 | 5 | 4 | 4 | 4 | 4 | 4 |
| | 32×32 | 2 | 4 | 1 | 4 | 4 | 5 | 4 | 5 | 5 | 4 | 4 |
| | 64×64 | 1 | 4 | - | 5 | 5 | 5 | 4 | 6 | 6 | 4 | 4 |
| =1 | 16×16 | 2 | 2 | - | - | - | 5 | - | - | - | 4 | 4 |
| | 32×32 | 2 | 4 | - | - | - | 5 | - | - | - | 4 | 4 |
| | 64×64 | 1 | 4 | - | - | - | 5 | - | - | - | 4 | 4 |
| U/V >1 | 16×16 | 2 | 2 | - | 2 | 2 | 4 | 3 | 3 | 3 | - | - |
| | 32×32 | 2 | 2 | 1 | 3 | 3 | 4 | 3 | 4 | 4 | - | - |
| | 64×64 | 1 | 3 | - | 4 | 4 | 4 | 3 | 5 | 5 | - | - |
| =1 | 16×16 | 2 | 2 | - | - | - | 4 | - | - | - | - | - |
| | 32×32 | 2 | 2 | - | - | - | 4 | - | - | - | - | - |
| | 64×64 | 1 | 3 | - | - | - | 4 | - | - | - | - | - |

a certain kernel, which are called as kernel bits, are presented in Table 2.

$R_Y$, $R_U$, and $R_V$ in Fig. 4 are all combined with the header bits and the kernel bits of each channel. For each channel, all of the quantized data of each parameter for a certain blocksize is encoded using the *Adaptive Arithmetic Coding* (AAC) [26].

Table 2 shows the details of bits allocation of a certain kernel towards different blocksizes. In Table 2, if a $B$ code $0$ or $10$ appears, there will be $NM$ models encoded as $n_P$ in Table 2. If a $B$ code $11$ appears, there will be four $16 \times 16$ blocks encoded as $(n_{NM} + n_{EG}) + n_P \times NM$, only the first of which requires the $B$ mark. For quantization of U and V channels, $\Sigma_{X_j Z_j}$ and $\Sigma_{Y_j Z_j}$ are set to be zero. Coding details of the parameters are discussed in the following sections.

*5.2.1. Eigenvalue decomposition of position matrix*

As for the $2 \times 2$ covariance matrix $\mathbf{R}_j$, two eigenvectors $\mathbf{V}_{2 \times 2} = \begin{bmatrix} V_{11} & V_{12} \\ V_{21} & V_{22} \end{bmatrix}$ and two eigenvalues $e_{1_j}$, $e_{2_j}$ are obtained through eigen-decomposition. According to [8], the covariance matrix can be implemented by encoding $e_{1_j}$, $e_{2_j}$, and $\eta_j$ which is the angle of principal eigenvector towards the positive X-axis.

Angle $\eta_j$ corresponds to the smaller eigenvalue between $e_{1_j}$ and $e_{2_j}$ denoted as $e_{a_j}$, so it is drawn from

$$\eta_j = \arctan \frac{V_{2a}}{V_{1a}}, \quad (30)$$

from which angle $\eta_j$ is known to be ranging from $-90°$ to $90°$ based on our sufficient experiments. Moreover, the prior $\alpha_j$ does not need to be encoded and is estimated by

$$\alpha_j = \left( \frac{1}{NM} + \frac{e_{1_j} e_{2_j}}{\sum_{i=1}^{NM} e_{1_i} e_{2_i}} \right)/2. \quad (31)$$

*5.2.2. Special case of a single-kernel block*

If a block is optimally represented by one model, it can be recovered equally well by using either Epanechnikov or Gaussian regressions. The reason lies in the particularity of the EM algorithm under the case of one model in Algorithm 1. After the initialization of mean value matrix $\boldsymbol{\mu}_1$ and covariance matrix $\boldsymbol{\Sigma}_1$ of the original image data, we can then obtain the result of posterior probability $Qt_{ij}$ (E-step) utilizing Equation (27) and (28). When the number of kernels $j = 1$, then $Qt_{i1} = 1$ and M-step turns out to be

$$\boldsymbol{\mu}_1 = \frac{\sum_{i=1}^{N} \boldsymbol{\varphi}_i}{N}, \quad \boldsymbol{\Sigma}_1 = \frac{\sum_{i=1}^{N} (\boldsymbol{\varphi}_i - \boldsymbol{\mu}_1)^T (\boldsymbol{\varphi}_i - \boldsymbol{\mu}_1)}{N}, \quad \alpha_1 = 1. \quad (32)$$

In this case, the iteration results are only related to the samples $\boldsymbol{\varphi}_i$, $i = 1, \ldots, N$. Therefore, we do not use any bit to tell the model type under such a case, and without loss of generality, we model the $16 \times 16$ blocks by EK while $32 \times 32$ and $64 \times 64$ by GK.

As for a certain block size, the 2-D position samples $\boldsymbol{\delta}_i$, $i = 1, \ldots, N$ are determined, thus the position mean value $\hat{\boldsymbol{\mu}}_j$ and position covariance matrix $\mathbf{R}_j$ are determined. Only the gray value $z_i$, $i = 1, \ldots, N$ are different. Therefore, only $\Sigma_{X_1 Z_1}$, $\Sigma_{Y_1 Z_1}$ and $\mu_{Z_1}$ are encoded as shown in Table 2.

*5.3. Parameter decoding*

At the decoder, parameters are approximately recovered from the encoded bitstream. With $e_{1_j}$ as an example, $k$ is the encoded

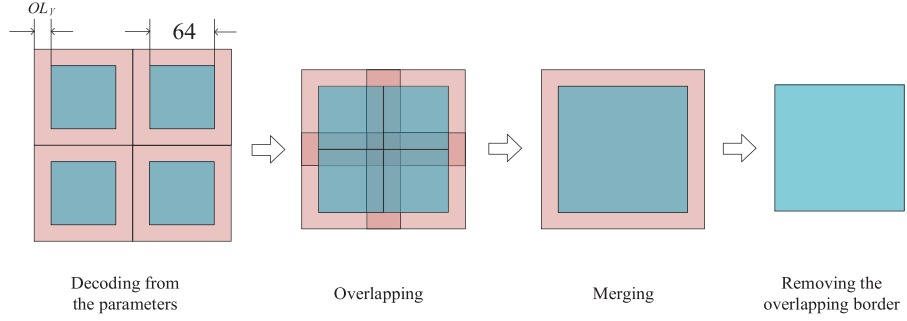

**Fig. 7.** An example of overlapping reconstruction of four 64×64 blocks of Y channel. The blue blocks represent the core contents and the pink borders represent the overlapping width. (For interpretation of the references to colour in this figure legend, the reader is referred to the web version of this article.)

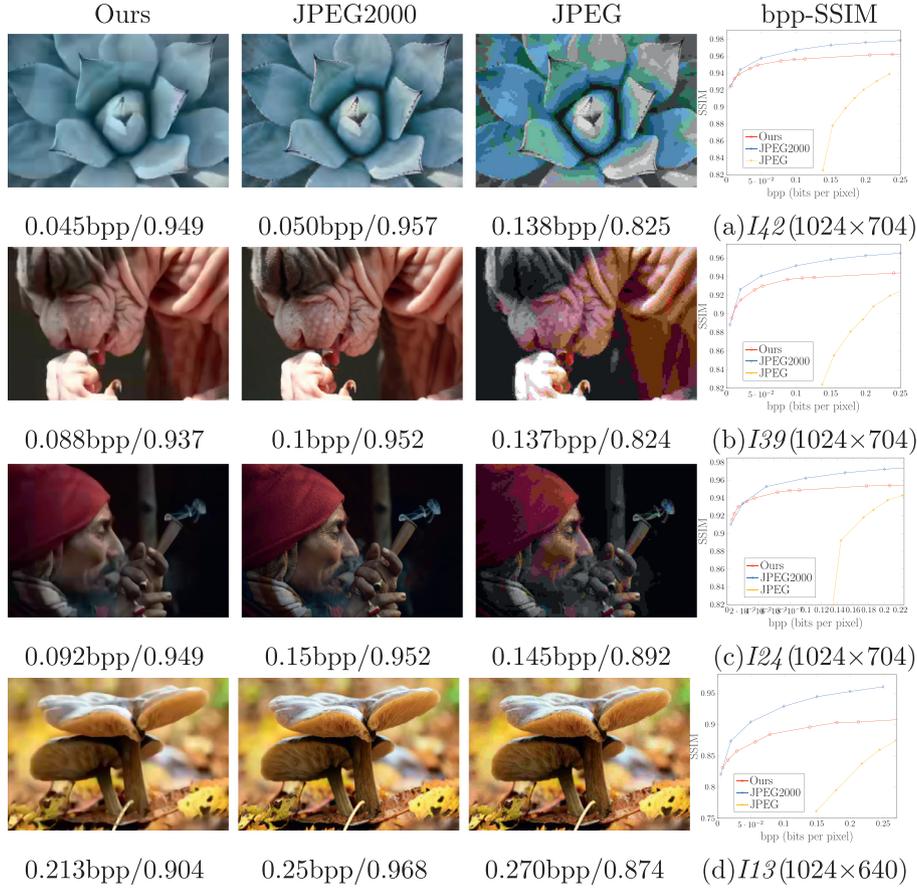

**Fig. 8.** Examples of our method, JPEG2000, and JPEG, together with the bpp-SSIM map. The bpp and SSIM of the chosen images are marked below.

bit representation of the original eigenvector $e_{1_j}$, which can be reconstructed as follows:

$$e_{1_j}[k] = M_{e_1} + (k-1) \cdot \frac{E_{e_1}}{2^{n_{e_{1_j}}} - 1}, k = 1, 2, \cdots, 2^{n_{e_{1_j}}}, \quad (33)$$

in which $M_{e_1}$ and $E_{e_1}$ represent the minimum and span of $e_{1_j}$. With all the decoded parameters, modeled image $I_p$ can be restored through (2).

After the images are reconstructed, the border merging is necessary. The implement of AMS has included the overlapping pixels; therefore, the reconstruction can realize a smoother transition from block to block. According to Fig. 7, the reconstruction is based on merging the overlapping border in proportion.

After the blocks are reconstructed, we implement a *Deblocking* algorithm to further weaken the block artifact. The deblocking is done for all the block borders by copying the boundary pixels and overlapping in proportion. For Y, the duplicate pixel width is 2 for 16×16 block. As for 32×32 block, it is 3 when $\lambda > 800$, 2 when $\lambda \leq 800$. As for 64×64 block, the width is 5, 3, and 2 for $\lambda > 800$, $100 < \lambda \leq 800$, and $\lambda \leq 100$. In addition to U and V, the width is 2 for 16×16 block and for 32×32 and 64×64 block, is respectively 3 and 6 when $\lambda > 200$, 2 and 4 when $\lambda \leq 200$. A *Block-matching and 3D filtering* (BM3D) algorithm [27] is finally added, with $\sigma = 15$ for Y and $\sigma = 20$ for U and V.

## 6. Experimental results

We compare the reconstructed results of some of the test images in case of specific bits with JPEG and JPEG2000 in Fig. 8 and Fig. 9. The test images of our experiments are derived from the *DIV2K dataset* [28]. In our algorithm, different bit consumption is realized by adjusting $\lambda$, which is chosen from 50000, 10000, 3200, 800, 400, 100, and 20 from low to high bit rates in our experiments. If $\lambda < 100$, we obtain the encoding results by increasing the bits of each parameter based on Table 2. As for the setting of overlapping border, if $\lambda > 3200$, $OL_Y=6$, $OL_{UV}=8$; if $400<\lambda \leq 3200$, $OL_Y=4$, $OL_{UV}=5$; if $100<\lambda \leq 400$, $OL_Y=2$, $OL_{UV}=3$; if $\lambda \leq 100$, $OL_Y=1$, $OL_{UV}=2$. The JPEG is implemented with QF=2, 5, 8, 10, 12, and 15. As for JPEG2000, the input bpps are setted at 0.005bpp, 0.02bpp, 0.05bpp, 0.1bpp, 0.15bpp, 0.2bpp, and 0.25bpp. All of the experiments using different algorithms are implemented using Matlab 2017b.

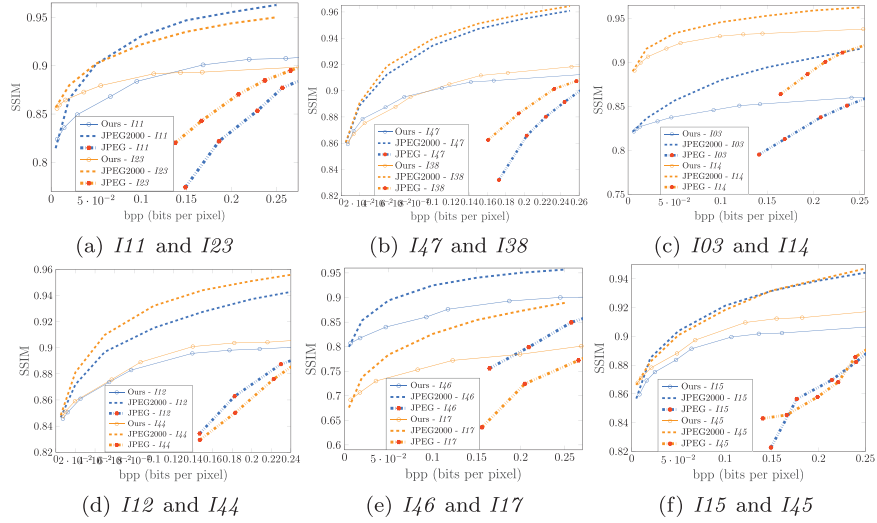

**Fig. 9.** Comparison bpp-SSIM maps of other images among ours, JPEG and JPEG2000. *I23* and *I38* are 1024×768, *I44* is 768×1024, and *I45* is 1024×576. Others are of 1024×704.

From the four test images in Fig. 8, we show the specific image results. The bpp-SSIM curves of other images are shown in Fig. 9.

The bpp-SSIM maps show that our method considerably outperforms JPEG and can also approach JPEG2000 at lower bitrates. At low bitrates, our method can achieve a competitive visual effect over JPEG below 0.25bpp with only 10% - 20% of bits consumed. Additionally, the SSIM is generally up to 0.85, which demonstrates superior compression performance for low reconstruction qualities, compared to JPEG. As the number of consumed bits increases, the effect of our method flattens out. From the images, it can be seen that our reconstructed images are smooth and color adaptive, which is suitable for human vision.

However, our framework is not pixel-based, which may bring a great property in low bitrates and has less block effect compared with JPEG, but the reconstruction is also lacking details in contrast to the JPEG2000. At high bitrates, the high-frequency information may become necessary, therefore our modeling-based method is less competitive compared with transform-based algorithms when the bpp becomes larger.

Regarding the consumption of time, our framework is implemented using an Intel(R) Core(TM) i7-7700K CPU @ 4.20GHz. The computational runtime for a single image is about one hour. A total of 16.7% of the time is spent on the U and V channels. The most time-consuming step is the K-means++ initialization.

## 7. Conclusion and future work

In conclusion, our EK and its correlated theory perform well in image coding especially in smaller blocks because the EK has a strong discontinuity, which enhances local support. This is also probably the reason that as the number of models increases, the close interaction between the models makes the EK more dominant than the GK. Our framework is an entirely modeling-based algorithm combined with EMR in smaller blocks, GMR in larger ones and a kernel selection is made for the middle-size blocks. Therefore, the proposed AMS algorithm can successfully leverage the two kernels. In essence, the framework enriches the practical significance of Epanechnikov kernel in image coding, which is an extension of Gaussian-only kernels in modeling-based compression.

The 3-D EK we deduced jumps out of the common usage of GK and provides more possibilities for applications of different models. The derivation process in this paper can be used as a reference for other kernel functions to be applied to regression modeling in the future. Our model-based image coding algorithm does not perform well at high bit rates, which can be optimized in the future. The optimization theories of EMR also need to be explored in order to anticipate with those new-developed SMoE algorithms. However, the goal of this paper is only to show the potential of using the EK in the application of image coding. We are looking forward to more applications for other fields based on Epanechnikov-correlated theories.

## Declaration of Competing Interest

The authors declare that they have no known competing financial interests or personal relationships that could have appeared to influence the work reported in this paper.

## Appendix A. Triple integral details of $\int\int_{\Omega_{\hat{\varphi}}}\int \left(1 - \hat{\varphi}^T \Lambda \hat{\varphi}\right) d\Omega_{\hat{\varphi}}, \left\{\Omega_{\hat{\varphi}} : \hat{\varphi}^T \Lambda \hat{\varphi} \leq 1\right\}$

In order to compute a triple integral, we need to convert the variable of standard coordinates $(\hat{x}, \hat{y}, \hat{z})$ to that of spherical coordinates $(r, \theta, \varphi), 0 \leq r \leq 1, 0 \leq \theta \leq 2\pi, 0 \leq \varphi \leq \pi$

$$\begin{cases} \hat{x} = ar\sin\varphi\cos\theta \\ \hat{y} = br\sin\varphi\sin\theta \\ \hat{z} = cr\cos\varphi \end{cases} \qquad (A.1)$$

and

$$J = \frac{\partial(\hat{x}, \hat{y}, \hat{z})}{\partial(r, \theta, \varphi)} = \begin{vmatrix} \frac{\partial \hat{x}}{\partial r} & \frac{\partial \hat{x}}{\partial \theta} & \frac{\partial \hat{x}}{\partial \varphi} \\ \frac{\partial \hat{y}}{\partial r} & \frac{\partial \hat{y}}{\partial \theta} & \frac{\partial \hat{y}}{\partial \varphi} \\ \frac{\partial \hat{z}}{\partial r} & \frac{\partial \hat{z}}{\partial \theta} & \frac{\partial \hat{z}}{\partial \varphi} \end{vmatrix} = abcr^2\sin\varphi. \qquad (A.2)$$

Thus, the integral turns out to be

$$\iiint_{\Omega_{\hat{\varphi}}:\hat{\varphi}^T \mathbf{\Lambda} \hat{\varphi} \leq 1} \left(1 - \hat{\varphi}^T \mathbf{\Lambda} \hat{\varphi}\right) d\Omega_{\hat{\varphi}}$$

$$= \iiint_{\frac{\hat{x}^2}{a^2}+\frac{\hat{y}^2}{b^2}+\frac{\hat{z}^2}{c^2} \leq 1} \left[1 - \left(\frac{\hat{x}^2}{a^2}+\frac{\hat{y}^2}{b^2}+\frac{\hat{z}^2}{c^2}\right)\right] d\hat{x} d\hat{y} d\hat{z}$$

$$= abc \int_0^{2\pi} d\theta \int_0^{\pi} d\varphi \int_0^1 \left(1 - r^2\right) r^2 \sin\varphi \, dr = \frac{8\pi abc}{15} = \frac{8\pi}{15|\mathbf{\Lambda}|}. \quad \text{(A.3)}$$

**Appendix B. Calculation of $\hat{\Sigma}$**

$\hat{X}$, $\hat{Y}$, and $\hat{Z}$ are the random value of $\hat{x}$, $\hat{y}$, and $\hat{z}$, therefore the covariance matrix of unit-sphere EK is

$$\hat{\Sigma} = \begin{bmatrix} \text{Cov}(\hat{X},\hat{X}) & \text{Cov}(\hat{X},\hat{Y}) & \text{Cov}(\hat{X},\hat{Z}) \\ \text{Cov}(\hat{Y},\hat{X}) & \text{Cov}(\hat{Y},\hat{Y}) & \text{Cov}(\hat{Y},\hat{Z}) \\ \text{Cov}(\hat{Z},\hat{X}) & \text{Cov}(\hat{Z},\hat{Y}) & \text{Cov}(\hat{Z},\hat{Z}) \end{bmatrix}. \quad \text{(B.1)}$$

We take the calculation of $\text{Cov}(\hat{X},\hat{X})$ as an example:

$$\text{Cov}(\hat{X},\hat{X}) = E(\hat{X}^2) - E^2(\hat{X}). \quad \text{(B.2)}$$

Because the distribution of $\hat{f}(\hat{\varphi})$ is even symmetric about the origin of the coordinate system. Firstly, we can get the result of $E(\hat{X})$ easily:

$$E(\hat{X}) = \iiint \hat{x} \cdot \hat{f}(\hat{\varphi}) d\hat{x} d\hat{y} d\hat{z}$$

$$= \iiint_{\frac{\hat{x}^2}{a^2}+\frac{\hat{y}^2}{b^2}+\frac{\hat{z}^2}{c^2} \leq 1} \hat{x} \cdot \frac{15}{8\pi}\left[1 - \left(\frac{\hat{x}^2}{a^2}+\frac{\hat{y}^2}{b^2}+\frac{\hat{z}^2}{c^2}\right)\right] d\hat{x} d\hat{y} d\hat{z} = 0 \quad \text{(B.3)}$$

Next, we can figure out $E(\hat{X}^2)$ according to (A.1) and (A.2).

$$E(\hat{X}^2)$$
$$= \iiint_{\frac{\hat{x}^2}{a^2}+\frac{\hat{y}^2}{b^2}+\frac{\hat{z}^2}{c^2} \leq 1} \hat{x}^2 \cdot \frac{15}{8\pi}\left[1 - \left(\frac{\hat{x}^2}{a^2}+\frac{\hat{y}^2}{b^2}+\frac{\hat{z}^2}{c^2}\right)\right] d\hat{x} d\hat{y} d\hat{z}$$
$$= \frac{15}{8\pi} \int_0^{2\pi} d\theta \int_0^{\pi} d\varphi \int_0^1 a^2(r\sin\varphi\cos\theta)^2(1-r^2)r^2\sin\varphi \, dr = \frac{a^2}{7}$$
$$\quad \text{(B.4)}$$

Thus, $\text{Cov}(\hat{X},\hat{X}) = a^2/7$. Similarly, we can know $\text{Cov}(\hat{Y},\hat{Y}) = b^2/7$, $\text{Cov}(\hat{Z},\hat{Z}) = c^2/7$, $\text{Cov}(\hat{X},\hat{Y}) = \text{Cov}(\hat{Y},\hat{X}) = 0$, $\text{Cov}(\hat{X},\hat{Z}) = \text{Cov}(\hat{Z},\hat{X}) = 0$, $\text{Cov}(\hat{Y},\hat{Z}) = \text{Cov}(\hat{Z},\hat{Y}) = 0$. Therefore,

$$\hat{\Sigma} = \frac{1}{7}\text{diag}(a^2, b^2, c^2) = \frac{1}{7}(\mathbf{\Lambda}^{-1})^2. \quad \text{(B.5)}$$

**Supplementary material**

Supplementary material associated with this article can be found, in the online version, at 10.1016/j.sigpro.2021.108090

**CRediT authorship contribution statement**

**Boning Liu:** Conceptualization, Methodology, Software, Validation, Formal analysis, Visualization, Writing - original draft, Investigation. **Yan Zhao:** Resources, Writing - review & editing, Project administration, Funding acquisition. **Xiaomeng Jiang:** Validation, Formal analysis, Investigation. **Shigang Wang:** Resources, Funding acquisition.